\documentclass[12pt,1p,review]{elsarticle}

\usepackage{graphicx}% Include figure files
\usepackage{dcolumn}% Align table columns on decimal point
\usepackage{bm}% bold math
\usepackage{amssymb}
\usepackage{comment}
\usepackage{subfigure}
\usepackage{tikz}
\usetikzlibrary{scopes}
\usetikzlibrary{calc}
\usepackage{amsthm}
\usepackage{amsopn}
\usepackage{amsmath}

\newcommand{\imag}{\rm{Im}}

\journal{Journal of Physics B}

\begin{document}

\title{Temperature-resolution anomalies in the reconstruction of time dynamics from energy-loss experiments}%

\begin{frontmatter}

\author{Anshul Kogar}
\author{Sean Vig}
\author{Yu Gan}
\author{Peter Abbamonte}
\address{Department of Physics and Federick Seitz Materials Research Laboratory, University of Illinois, Urbana, IL, 61801}
\address{Advanced Photon Source, Argonne National Laboratory, Argonne, IL, 60439}

\begin{abstract}
Inelastic scattering techniques provide a powerful approach to studying electron and nuclear dynamics, via reconstruction of a propagator that quantifies the time evolution of a system. There is now growing interest in applying such methods to very low energy excitations, such as lattice vibrations, but in this limit the cross section is no longer proportional to a propagator. Significant deviations occur due to the finite temperature Bose statistics of the excitations. Here we consider this issue in the context of high-resolution electron energy loss experiments on the copper-oxide superconductor Bi$_2$Sr$_2$CaCu$_2$O$_{8+x}$. We find that simple division of a Bose factor yields an accurate propagator on energy scales greater than the resolution width. However, at low energy scales, the effects of resolution and finite temperature conspire to create anomalies in the dynamics at long times. We compare two practical ways for dealing with such anomalies, and discuss the range of validity of the technique in light of this comparison.
\end{abstract}


\end{frontmatter}

\section{Introduction}
\label{sec:intro}
Electron dynamics underlie all fundamental phenomena in chemical and materials physics, requiring the use of fast, time-resolved techniques to probe and understand their fundamental properties. Studies of fast dynamics may be carried out in two distinct ways. The most common, which one might call the ``explicit approach," is to study time dynamics with an ultrafast source, the time dependence being accessed by varying a controlled delay between pump and probe pulses. A much less common approach is to perform momentum-resolved scattering experiments, such as inelastic x-ray, electron, or neutron scattering, and to use inverse methods to reconstruct the dynamics indirectly\cite{water,GenInt}. The main advantage of the latter is that it provides better time resolution as well as quantitative imaging, which is preferable for many applications. This approach has been used, for example, to image exciton dynamics in large-gap insulators\cite{LiF}, solvation phenomena in liquid water\cite{coridan}, and to measure the effective fine structure constant of graphene\cite{reed}. 

What makes the energy-domain approach possible is that the cross section for inelastic scattering experiments is related to the imaginary part of a response function with well-defined causal properties. For the case of x-rays, for example, this quantity is the density response function, $\chi(q,\omega)$, which describes how a disturbance in the density propagates about the system. Its causal character allows the excitations to be properly ordered in time by use of the Kramers-Kronig relations. 

In the case of very high energy resolution experiments, a subtlety arises that so far has not been addressed. The cross section is not, strictly speaking, proportional to the response function itself, but to a correlation function, $S(q,\omega)$. The two quantities are related by the fluctuation-dissipation theorem, which says that $S$ is proportional to $\imag [\chi]$, the proportionality constant being a Bose factor that describes the quantum statistics of the excitations. The validity of this framework is predicated on the presence of thermodynamic equilibrium and the equipartition of energy. 

At high energy scales, i.e., $\omega \gg k_B T$, the Bose factor is equal to unity (for positive $\omega$) and it is unnecessary to draw a distinction between $S$ and $\imag[\chi]$. For very high resolution experiments, however, the two quantities are qualitatively different and corrections for the Bose factor must be made for the time dynamics to be refined. 

In this paper we investigate this issue in the context of high-resolution electron energy loss spectroscopy (HR-EELS), which is an inelastic electron scattering technique that can achieve energy resolution close to 1 meV--far below the $k_B T \sim 25 meV$ at room temperature. Our key finding is that, for modest energy scales, the Bose factor may simply be divided out of the cross section, resulting in a causal response function with the proper symmetry properties. At low energy scales, however, the effects of finite temperature and finite spectrometer resolution conspire to create anomalies in the dynamics on very long time scales. We illustrate this effect with some HR-EELS measurements on the high temperature superconductor Bi$_2$Sr$_2$CaCu$_2$O$_{8+x}$.

This manuscript is organized as follows. In Section 2 we derive the HR-EELS cross section from basic principles, approximately following the work of Mills and Ibach from the 1970s \cite{IbachMills,mills1972}. In Section 3, we identify the relevant correlation and response functions, and write down the exact form of the associated fluctuation dissipation theorem. Most of the content of these two sections may be found in various locations in the literature from forty years ago. Our purpose here is to synthesize the essential elements using notation that is standard in modern condensed matter physics, which is convenient for discussing the inverse problem. In Section 4 we present measurements from Bi$_2$Sr$_2$CaCu$_2$O$_{8+x}$, from which we illustrate the temperature-resolution anomaly. We close by discussing the range of validity of various inelastic scattering techniques in light of this effect.

\section{HR-EELS Cross Section}

The contents of this section were synthesized from Refs. \cite{IbachMills,mills1972,mills1975,mills1980,persson}. We begin by revisiting the cross section for HR-EELS, using contemporary notation that is preferable for discussing the inverse problem. In HR-EELS, a probe electron with well defined energy and momentum, ${\bf k}_1$ and $\omega_1$, is incident on a sample surface, and scatters to some final state ${\bf k}_2$ and $\omega_2$. To determine the cross section for this process, we wish to determine the matrix element

\begin{equation}
M =  - \frac{i}{\hbar }\left\langle f \right|H'(0)\left| i \right\rangle
\end{equation}

\noindent
For the case of HR-EELS, the relevant interaction is the instantaneous Coulomb interaction, 

\begin{equation}
H'=\frac{e^2}{2} \int{ \frac{\hat{\rho}({\bf R}_1) \hat{\rho}({\bf R}_2)}{\left | {\bf R}_1-{\bf R}_2 \right | }  }
\end{equation}

\noindent
where $\hat{\rho}$ is the electron number density operator and the coordinate ${\bf R} = ({\bf r},z)$, where ${\bf r}$ and $z$ are components parallel and perpendicular to the surface, respectively. In terms of this interaction, the scattering matrix element is given by

\begin{equation}
M=\frac{e^2}{2} \int{ \frac{ \left\langle n \right| \hat{\rho}({\bf R}_1) \left | m \right > \psi_s^*({\bf R}_2) \psi_i({\bf R}_2)}{\left | {\bf R}_1-{\bf R}_2 \right | } }
\end{equation}

\noindent
where $\psi_i$ and $\psi_s$ are the wave functions for the initial and final state of the probe electron, and $\left | m \right >$ denotes a many-body eigenstate of the semi-infinite material system. Following past practice in the HR-EELS field\cite{IbachMills,mills1972}, in Eq. 3 we have neglected exchange scattering, which can be important if the overlap between the probe and valence electron wave functions is significant\cite{MillsSpin}. In so doing we have neglected the possibility of spin-dependent scattering, which can be significant in materials exhibiting pronounced magnetic excitations, such as magnons.

In HR-EELS, multiple scattering effects are significant. One of the crucial milestones for the technique was the recognition that multiple scattering predominantly takes place in the elastic, rather than the inelastic, channel. This suggests that the scattering can be accurately described by using wave functions for the probe electron that are modified from their nominally plane-wave form, and treating the inelastic scattering in the Born approximation. As was argued earlier by Mills \cite{mills1972}, the appropriate incident and scattered wave functions, $\psi_i$ and $\psi_s$, are

\begin{equation}
\begin{array}{rcl}
{\psi _i}({\bf{R}}) &=& {N_i}\left( {{e^{i{{\bf{k}}_i} \cdot {\bf{r}}}}{e^{ik_i^zz}} + {R_i}{e^{i{{\bf{k}}_i} \cdot {\bf{r}}}}{e^{ - ik_i^zz}}} \right)\theta (z)
\\
{\psi _s}({\bf{R}}) &=& {N_s}\left( {{e^{i{{\bf{k}}_s} \cdot {\bf{r}}}}{e^{ik_s^zz}} + {R_s}{e^{i{{\bf{k}}_s} \cdot {\bf{r}}}}{e^{ - ik_s^zz}}} \right)\theta (z)
\end{array}
\end{equation}

\noindent
In this expression, $R_i$ and $R_s$ describe the effect of specular reflection of the incident or scattered plane wave off the sample surface, and the step function, $\theta(z)$, implies that the wave functions do not penetrate into the material, which we take to fill the half-space $z<0$. $N_i$ and $N_s$ are normalization constants that, if the phase shift due to the reflection is small, have the form\cite{IbachMills,mills1972}

\begin{equation}
N_{i,s}^{} = \sqrt {\frac{2}{{V\left( {1 + {{\left| {{R_{i,s}}} \right|}^2}} \right)}}}.
\end{equation}

Inserting these expressions into Eq. 3 results in four distinct terms that contribute to the inelastic scattering cross section. As was shown previously by Mills \cite{mills1972}, the matrix element is dominated by the cross terms, which involve single powers of $R_i$ and $R_s$. Keeping only these two terms, the matrix element is given by

\begin{equation}
M = {M_i} + {M_s}
\end{equation}

\noindent
where 

\begin{equation}
{M_{i,s}} =  - \frac{{i{e^2}}}{{2\hbar }}N{R_{i,s}} \int{ {\frac{{\left\langle n \right|\hat \rho ({{\bf{R}}_1})\left| m \right\rangle {e^{i{\bf{q}} \cdot {{\bf{r}}_2}}}{e^{\mp i\left( {k_s^z + k_i^z} \right){z_2}}}\theta ({z_2})}}{{\left| {{{\bf{R}}_1} - {{\bf{R}}_2}} \right|}}d{{\bf{R}}_1}d{{\bf{R}}_2}} }
\end{equation}

\noindent
where $N=\sqrt{N_i N_s}$ and ${\bf q}$ is the in-plane component of the momentum transfer. Expressed explicitly in terms of in- and out-of-plane coordinates,

\begin{equation}
{M_{i,s}} =  - \frac{{i{e^2}}}{{2\hbar }}{N^2}{R_{i,s}} \int{ {\frac{{\left\langle n \right|\hat \rho ({{\bf{r}}_1},{z_1})\left| m \right\rangle {e^{i{\bf{q}} \cdot {{\bf{r}}_2}}}{e^{\mp i\left( {k_s^z + k_i^z} \right){z_2}}}\theta ({z_2})}}{{\sqrt {{{\left( {{r_1} - {r_2}} \right)}^2} + {{\left( {{z_1} - {z_2}} \right)}^2}} }}d{\bf{r}}_1^2d{\bf{r}}_2^2d{z_1}d{z_2}} } .
\end{equation}

\noindent
We begin by considering $M_s$ only. Performing the $r_2$ integral yields

\begin{equation}
{M_s} =  - \frac{i}{{2\hbar }}{N^2}{R_s}\,{V_{2D}}(q) \int{ {d{\bf{r}}_1^2d{z_1}d{z_2}\left\langle n \right|\hat \rho ({{\bf{r}}_1},{z_1})\left| m \right\rangle {e^{i\left( {k_s^z + k_i^z} \right){z_2}}}} \theta ({z_2})\,{e^{i{\bf{q}} \cdot {{\bf{r}}_1}}}\,{e^{ - q\left| {{z_1} - {z_2}} \right|}} }
\end{equation}

\noindent
where $q = |{\bf q}|$ and $V_{2D}(q)=2\pi e^2/q$ is the propagator for the Coulomb interaction in two dimensions. The ${\bf r}_1$ integral is just a Fourier transform, so

\begin{equation}
{M_s} =  - \frac{i}{{2\hbar }}{N^2}{R_s}\,{V_{2D}}(q) \int_{-\infty}^0 {d{z_1}\left\langle n \right|\hat \rho \left( {{\bf{q}},{z_1}} \right)\left| m \right\rangle } \int_0^{\infty} {d{z_2}{e^{i\left( {k_s^z + k_i^z} \right){z_2}}}{e^{ - q\left| {{z_1} - {z_2}} \right|}}},
\end{equation}

\noindent
where we have used the fact that the material is semi-infinite, i.e., the $z_1$ integrand is nonzero only for $z_1<0$. The quantity $(z_1-z_2)$ is always negative, so the $z_2$ integral may readily be done to yield

\begin{equation}
{M_s} = \frac{{{N^2}{R_s}}}{{2\hbar }}\frac{{{V_{2D}}(q)}}{{k_s^z + k_i^z + iq}} \int_{-\infty}^0 {\left\langle n \right|\hat \rho \left( {{\bf{q}},{z_1}} \right)\left| m \right\rangle {e^{q{z_1}}}}.
\end{equation}

In this form, it is clear why the cross terms, Eq. 6, dominate the scattering cross section. In high energy, bulk-sensitive EELS, the inelastic cross section $\sim 1/q^4$, i.e., is a rapidly decreasing function of $q$. In cross terms like Eq. 11, however, the denominator contains the sum  $k_s^z + k_i^z$, rather than the difference $q^z = k_s^z - k_i^z$,  $q^z$ being the out-of-plane component of the momentum transfer. Hence, in the so-called ``dipole" regime, in which measurements are carried out in near-specular geomtery, $k_s^z \approx -k_i^z$,  so the sum approximately vanishes and only the in-plane component, $q$, appears in the denominator. The overall effect is that the probe electron undergoes a large change in its out-of-plane momentum, but this transferred momentum comes ``for free," in the sense that it does not enter the Coulomb propagator, the momentum being supplied by the reflectance from the sample surface, rather than the inelastic event. 

The other part of the matrix element, $M_i$, is identical to the above, with $R_s \rightarrow R_i$ and $k_s^z + k_i^z \rightarrow -k_s^z - k_i^z$. This gives

\begin{equation}
{M_i} = \frac{{{N^2}{R_i}}}{{2\hbar }}\frac{{{V_{2D}}(q)}}{{ - k_s^z - k_i^z + iq}} \int_{-\infty}^0 {d{z_1}\left\langle n \right|\hat \rho \left( {{\bf{q}},{z_1}} \right)\left| m \right\rangle {e^{q{z_1}}}}.
\end{equation}

\noindent
The full matrix element $M=M_i+M_s$ is then

\begin{equation}
M = \frac{{ - i}}{\hbar }{N^2}R\frac{{4\pi {e^2}}}{{{{\left( {k_s^z + k_i^z} \right)}^2} + {q^2}}} \int_{-\infty}^0 {d{z_1}\left\langle n \right|\hat \rho \left( {{\bf{q}},{z_1}} \right)\left| m \right\rangle {e^{q{z_1}}}}
\end{equation}

\noindent where for simplicity we have assumed $R_s=R_i=R$. 

To turn this matrix element into a scattering cross section we take the traditional route of applying Fermi's Golden Rule. The double differential scattering cross section is defined as

\begin{equation}
\frac{{{\partial ^2}\sigma }}{{\partial \Omega \partial E}} = \frac{1}{\Phi }\sum\limits_f {{w_{f \leftarrow i}}}  \cdot \frac{{{\partial ^2}N}}{{\partial \Omega \partial E}}
\end{equation}

\noindent
where $\Phi$ is the incident flux and $\partial^2 N / \partial \Omega \partial E $ is the density of final states of the scattered electron. The matrix element enters in the transition rate

\begin{equation}
{w_{f \leftarrow i}} = \frac{{2\pi }}{\hbar }{\left| {\left\langle f \right|H'(0)\left| i \right\rangle } \right|^2} = 2\pi \hbar \,{\left| M \right|^2}.
\end{equation}

For a single, nonrelativistic electron traveling at velocity $v$, $\Phi = v/V = \sqrt{2E_i/m}/V$, where $E_i$ is the incident electron kinetic energy and $V$ is the volume of all space. The density of final states is given by the usual expression 

\begin{equation}
\frac{{{\partial ^2}N}}{{\partial \Omega \partial E}} = \frac{V}{{8{\pi ^3}}}{\left( {\frac{{2m}}{{{\hbar ^2}}}} \right)^{3/2}}\sqrt E  .
\end{equation}

\noindent
Squaring the matrix element, the final result is

\begin{align}
\frac{{{\partial ^2}\sigma }}{{\partial \Omega \partial E}} &= {\sigma _0}{\left[ { V_{eff}(k_i^z,k_s^z,q)} \right]^2} 
\int_{-\infty}^0 {d{z_1}d{z_2}{e^{q\left( {{z_1} + {z_2}} \right)}}} \nonumber \\
& \times {\sum\limits_{m,n} {\left\langle n \right|\hat \rho \left( {{\bf{q}},{z_1}} \right)\left| m \right\rangle \left\langle m \right|\hat \rho \left( { - {\bf{q}},{z_2}} \right)\left| n \right\rangle } } \nonumber \\
&\times {P_m}\delta \left( {E - {E_n} + {E_m}} \right)
\end{align}

\noindent
where

\begin{equation}
{\sigma _0} \equiv \sqrt {\frac{{{E_f}}}{{{E_i}}}} \frac{{{m^2}}}{{2{\pi ^2}{\hbar ^4}}}\frac{{{{\left| R \right|}^2}}}{{{{\left( {1 + {{\left| R \right|}^2}} \right)}^2}}}
\end{equation}

\noindent and 

\begin{equation}
V_{eff}(k_i^z,k_s^z,q) \equiv \frac{{4\pi {e^2}}}{{{{\left( {k_s^z + k_i^z} \right)}^2} + {q^2}}}
\end{equation}

\noindent is an effective Coulomb propagator that describes how the probe electron couples to excitations near the surface of a semi-infinite system. From this result we can already confirm the crucial observation, made previously \cite{mills1972}, that the probe depth in HR-EELS is not set by the penetration depth of the electrons, as it is in other electron spectroscopies such as angle-resolved photoemission (ARPES) or scanning tunneling microscopy (STM), but by the inverse of the in-plane component of the momentum transfer, $q$. The reason is that HR-EELS measures the dielectric response of the surface, which is coupled electromagnetically to layers deeper in the material. Hence, at low $q$, the technique can couple to features “deep” in the sample, via their influence on the dielectric response near the surface.

\section{Fluctuation-Dissipation Theorem for HR-EELS}

We are now ready to establish a relationship between the cross section, Eq. 17, a correlation function for the density, and a density response function. In complete generality, for a many-body system, the density correlation function is defined as \cite{fetter}

\begin{equation}
S({{\bf{R}}_1},{{\bf{R}}_2},\omega ) = \frac{1}{\hbar} \sum\limits_{m,n} {\left\langle m \right|\hat \rho \left( {{{\bf{R}}_1}} \right)\left| n \right\rangle \left\langle n \right|\hat \rho \left( {{{\bf{R}}_2}} \right)\left| m \right\rangle } {P_m}\delta \left( {\omega - {\omega_n} + {\omega_m}} \right)
\end{equation}

\noindent in real space. This quantity can also be expressed in momentum space,

\begin{equation}
S({{\bf{Q}}_1},{{\bf{Q}}_2},\omega ) = \frac{1}{\hbar} \sum\limits_{m,n} {\left\langle m \right|\hat \rho \left( {{{\bf{Q}}_1}} \right)\left| n \right\rangle \left\langle n \right|\hat \rho \left( { - {{\bf{Q}}_2}} \right)\left| m \right\rangle } {P_m}\delta \left( {\omega - {\omega_n} + {\omega_m}} \right)
\end{equation}

\noindent where we have adopted the momentum notation ${\bf Q} = ({\bf q},q_z)$, where ${\bf q}$ and $q_z$ are the in-plane and out-of-plane components, respectively. For relation to the cross section, we consider here the mixed representation, 

\begin{equation}
S({{\bf{q}}_1},{z_1};{{\bf{q}}_2},{z_2};\omega ) = \frac{1}{\hbar} \sum\limits_{m,n} {\left\langle m \right|\hat \rho \left( {{{\bf{q}}_1},{z_1}} \right)\left| n \right\rangle \left\langle n \right|\hat \rho \left( { - {{\bf{q}}_2},{z_2}} \right)\left| m \right\rangle } {P_m}\delta \left( {\omega - {\omega_n} + {\omega_m}} \right) .
\end{equation}

\noindent In terms of this quantity, Eq. 17 may be written,

\begin{equation}
\frac{{{\partial ^2}\sigma }}{{\partial \Omega \partial E}} = {\sigma _0}{\left[ {V_{eff}(k_i^z,k_s^z,q)} \right]^2} \int_{-\infty}^0 
{d{z_1}d{z_2}{e^{q\left( {{z_1} + {z_2}} \right)}}S({\bf{q}},{z_1};{\bf{q}},{z_2};\omega )}
\end{equation}

\noindent This confirms the notion that, HR-EELS directly measures a correlation function for the electron density in the region near the surface of the material \cite{IbachMills}. To complete our study, we must identify a relationship between this quantity and a causal response function.

The the density response function is defined as \cite{fetter}

\begin{equation}
\chi ({{\bf{R}}_1},{{\bf{R}}_2};{t_1} - {t_2}) =  - \frac{i}{\hbar }\sum\limits_m {{P_m}\left\langle m \right|\left[ {\hat \rho ({{\bf{R}}_1},{t_1}),\hat \rho ({{\bf{R}}_2},{t_2})} \right]\left| m \right\rangle \,\theta ({t_1} - {t_2})}
\end{equation}

\noindent where $[,]$ represents a commutator. In contrast to the correlation function, $\chi ({\bf R}_1,{\bf R}_2;t_1 - t_2)$ is a propagator for the charge density, i.e., it represents the amplitude that a disturbance in the density at location ${\bf R}_2$ will propagate to ${\bf R}_1$ after elapsed time $t_1-t_2$. In contrast to $S$, $\chi$ is a microscopic representation of the collective charge dynamics of the system, exhibiting causality enforced by the $\theta ({t_1} - {t_2})$ term, which mandates that disturbances in the density can only influce the state of the system at later times. Written out explicitly in the mixed representation, the response function has the form

\begin{align}
 \chi ({{\bf{q}}_1},{z_1};{{\bf{q}}_2},{z_2},\omega ) &= \frac{1}{\hbar}  \left [ {\frac{{\left\langle m \right|\hat \rho ({{\bf{q}}_1},{z_1})\left| n \right\rangle \left\langle n \right|\hat \rho ({{\bf{q}}_2},{z_2})\left| m \right\rangle }}{{\omega  - {\omega _n} + {\omega _m} + i\eta }}} \right.\nonumber\\
 &\qquad \left. {} { - \frac{{\left\langle m \right|\hat \rho ({{\bf{q}}_1},{z_1})\left| n \right\rangle \left\langle n \right|\hat \rho ({{\bf{q}}_2},{z_2})\left| m \right\rangle }}{{\omega  + {\omega _n} - {\omega _m} + i\eta }}} \right ].
\end{align}

\noindent To relate this quantity to the correlation function, we begin by taking its imaginary part. Using the relation

\begin{equation}
{\mathop{\rm Im}\nolimits} \left[ {\frac{1}{{x + i\eta }}} \right] =  - \pi \,\delta (x)
\end{equation}

\noindent for infinitesimal $\eta$ we get

\begin{align}
&{\mathop{\rm Im}\nolimits} \left[ {\chi ({\bf{q}},{z_1}; - {\bf{q}},{z_2},\omega )} \right] = - \frac{{ \pi }}{\hbar } \sum_{m,n} P_m      \nonumber \\
& \times \left [ \left\langle m \right|\hat \rho ({\bf{q}},{z_1})\left| n \right\rangle \left\langle n \right|\hat \rho ( - {\bf{q}},{z_2})\left| m \right\rangle \right . \delta \left( {\omega  - {\omega _n} + {\omega _m}} \right) \nonumber \\
&- \left . \left\langle m \right|\hat \rho ({\bf{q}},{z_1})\left| n \right\rangle \left\langle n \right|\hat \rho ( - {\bf{q}},{z_2})\left| m \right\rangle \delta \left( {\omega  + {\omega _n} - {\omega _m}} \right) \right ]
 \end{align}

\noindent where, anticipating a comparison to Eq. 22, we have chosen the specific case ${\bf q}_1 = -{\bf q}_2 = {\bf q}$. The first term is identical to the correlation function,

\begin{align}
&{\mathop{\rm Im}\nolimits} \left[ {\chi ({\bf{q}},{z_1}; - {\bf{q}},{z_2},\omega )} \right] =  - \pi S({\bf{q}},{z_1};{\bf{q}},{z_2};\omega ) \nonumber \\
&+ \frac{\pi }{\hbar }\sum\limits_{m,n} {{P_m}\left[ {\left\langle m \right|\hat \rho ({\bf{q}},{z_1})\left| n \right\rangle \left\langle n \right|\hat \rho ( - {\bf{q}},{z_2})\left| m \right\rangle \delta \left( {\omega  + {\omega _n} - {\omega _m}} \right)} \right]} .
\end{align}

\noindent To deal with the second term, we recognize that 

\begin{equation}
{P_n} = {P_m}{e^{ - \beta \left( {{E_n} - {E_m}} \right)}} = {P_m}{e^{ - \beta \hbar \omega }}.
\end{equation}

\noindent Substituting this into Eq. 28 and exchanging the indices $m$ and $n$ gives

\begin{equation}
{\mathop{\rm Im}\nolimits} \left[ {\chi ({\bf{q}},{z_1}; - {\bf{q}},{z_2},\omega )} \right] =  - \pi S({\bf{q}},{z_1};{\bf{q}},{z_2};\omega ) + \pi {e^{\beta \hbar \omega }}S( - {\bf{q}},{z_1}; - {\bf{q}},{z_2};\omega )
\end{equation}

\noindent or, equivalently, 

\begin{equation}
S({\bf{q}},{z_1};{\bf{q}},{z_2};\omega ) =  - \frac{1}{\pi }\frac{1}{{1 - {e^{\beta \hbar \omega }}}}{\mathop{\rm Im}\nolimits} \left[ {\chi ({\bf{q}},{z_1}; - {\bf{q}},{z_2},\omega )} \right].
\end{equation}

\noindent Eq. 31 is a statement of the fluctuation-dissipation theorem relevant to high-resolution, electron energy loss spectroscopy. Its physical meaning is that the scattered intensity, which is directly proportional to the correlation function, $S$, is also a measure of the dissipative, imaginary part of the Green's function that describes the charge dynamics. The only assumption underlying this relationship is the presence of thermodynamic equilibrium and equipartition of energy. The proportionality factor, $n(\omega)=(1-\exp{\beta \hbar \omega})^{-1}$, is the so-called Bose factor, which mandates that the excitations that contribute to $S$ exhibit Bose statistics, which is required for a two-particle response function.

\section{Resolution Anomalies}

The relation of the HR-EELS cross section to a response function suggests that this method, when combined with inverse methods such as those described in Refs. \cite{GenInt,water}, could be used to image the dynamics of electrons near surfaces.

As discussed earlier, however, HR-EELS has very high energy resolution, and in the low-energy loss region will be sensitive to the non-trivial frequency region of the Bose factor, where there is a substantial difference between the correlation function, $S(q,\omega)$, and the response function $\imag[\chi(q,\omega)]$. To extract the latter, the Bose factor must be divided from the experimental data. 

In doing so, we encounter a problem. The cross section is proportional to $n(\omega) \imag[\chi(q,\omega)]$, but the product is convolved with the resolution function of the spectrometer. On energy scales similar to the width of this function, $n(\omega)$ cannot be divided out, impeding the reconstruction of the time dynamics.

\begin{figure}\centering
\includegraphics[width=0.9\textwidth]{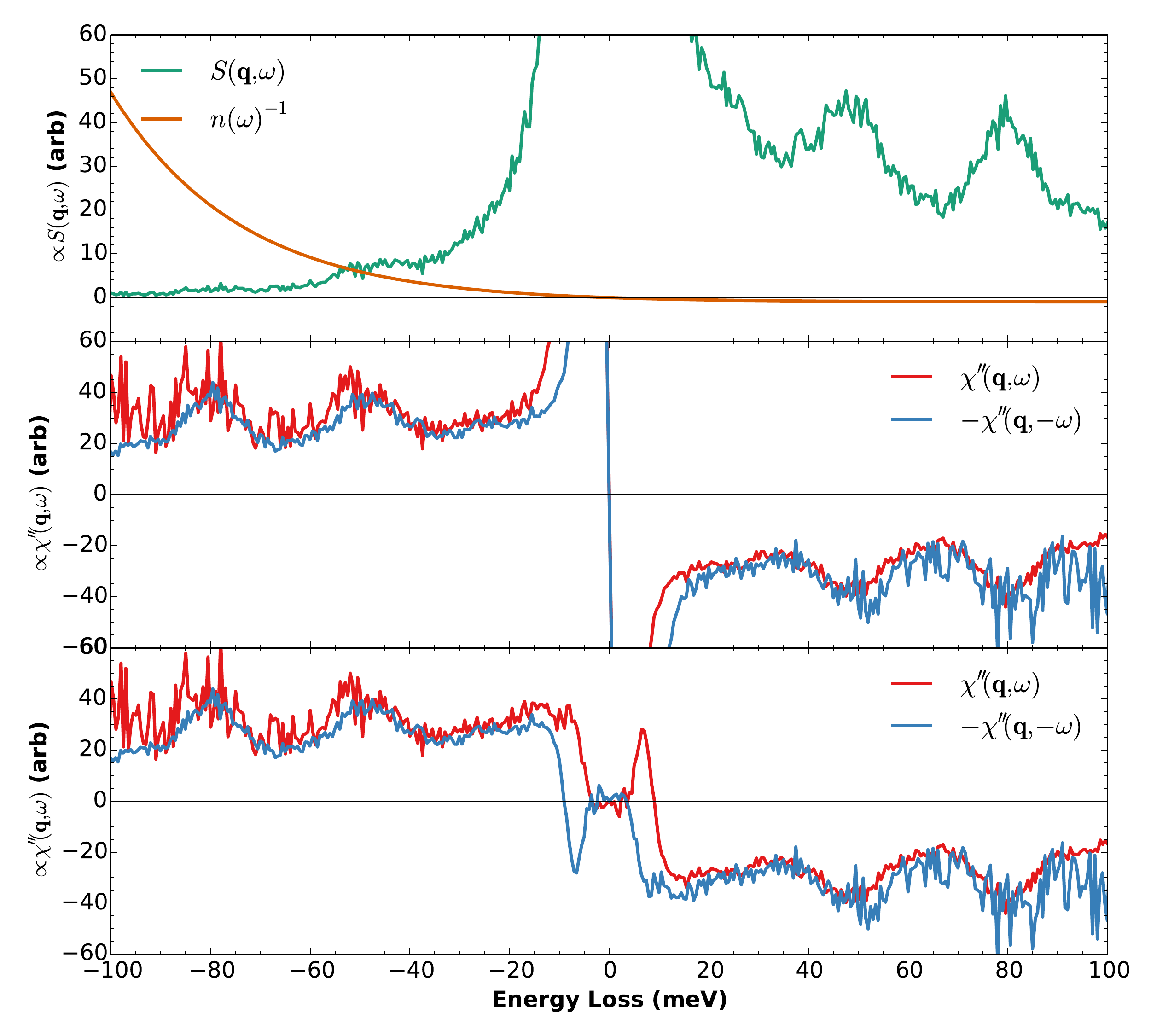}
\caption{Division of the Bose factor, $n(\omega)$, from HR-EELS data. (a) Raw HR-EELS data from the high temperature superconductor Bi$_2$Sr$_2$CaCu$_2$O$_{8+x}$, showing several collective modes that are visible in both the Stokes and the Antistokes regions. The inverse of the Bose factor, $n(\omega)^{-1}$ (orange line), is shown for comparison. (b) Result of the division of $n(\omega)$ from the data in part (a). The result is antisymmetric, as expected, except in a region near the zero-loss line where the symmetry is lost and a divergence is observed. (c) The same curve as in (b), but with the zero-loss feature first removed with a fit of a Voigt function. The divergence at $\omega=0$ is eliminated, but the antisymmetry is not perfectly recovered.
}
\end{figure}

To illustrate this point, we show in Fig. 1a an HR-EELS scan taken in the dipole regime from the copper-oxide superconductor Bi$_2$Sr$_2$CaCu$_2$O$_{8+x}$. The energy resolution for this scan was set to around 6 meV. Several collective modes are visible, which have been described previously by multiple authors\cite{persson1990, demuth1990, lieber1992, kesmodel1993, plummer2010}. In accordance with Eq. 23, both stokes and antistokes features are visible corresponding to the probe electron creating or annihilating an excitation, respectively, with a weight described by $n(\omega)$.

In contrast to $S$, we expect the dissipative part of response function to be odd in frequency, i.e., $\imag[\chi(q,\omega)]=-\imag[\chi(q,-\omega)]$. The reason is that the charge density in real space and time must be real, so $\chi^*(q,\omega)=\chi(q,-\omega)$\cite{martin}. Hence, we expect that dividing Fig. 1a by $n(\omega)$ should yield a function that is antisymmetric in $\omega$.

\begin{figure}\centering
\includegraphics[width=0.9\textwidth]{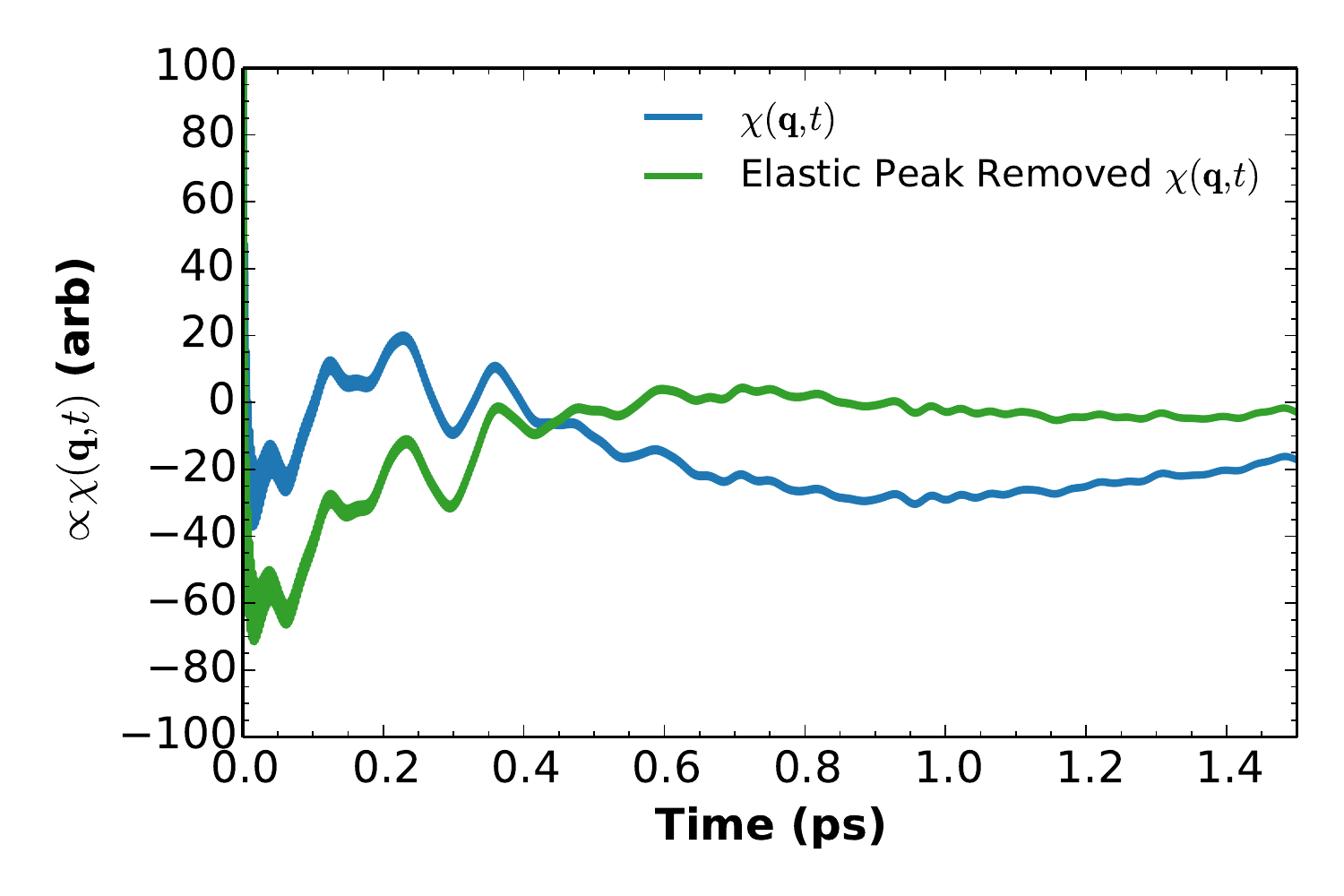}
\caption{Time dependence of the spatial Fourier component, $\chi(q,t)$, determined by taking a sine transform of each of the two spectra presented in Figs. 1b,c. The curves differ in a slowly varying background, though the short-time behavior is very similar.
}
\end{figure}

The result of the division is shown in Fig. 1b. At energy scales larger than the energy resolution, the division works remarkably well, with energy loss and gain features overlapping to within the experimental statistics. In the region near the elastic line, however, the symmetry is lost, and a divergence occurs at $\omega=0$. The divergence results from a rounding off of $S(q,\omega)$ near zero energy due to the finite resolution, and represents the inability to distinguish between truly static correlations and finite frequency excitations lying within the elastic line. Customarily, such features are collectively referred to as ``quasielastic scattering."

Because of this problem, an approach that is widespread in the literature (see, for example, Refs. \cite{water,larson1997}) is to fit the zero-loss line with a function and subtract it from the raw data. The argument for this approach is that $\imag[\chi(q,\omega)]$ should go to zero as $\omega \rightarrow 0$, so it is appropriate to start by eliminating the features at zero frequency. For comparison, we have taken this approach also, subtracting off a Voigt function fit to the elastic line. The result, after dividing out $n(\omega)$, is shown in Fig. 1c. The divergence observed previously has now been eliminated, but the result still is not properly antisymmetric in the low-energy region, indicating that such a fit has removed some relevant inelastic features. We are forced to conclude that there is no seamless way to eliminate resolution anomalies completely, and the best way forward is to characterize their effect on the dynamics to identify the range of validity of the approach. 

To get a feel for the effect of such anomalies, we transform both curves into the time domain and compare the two. While neither curve will be a perfect representation of the dynamics, comparing them is a good way to determine the range over which errors from the resolution anomaly have an influence. Transformation to space and time is achieved by performing a frequency sine transform of $\imag[\chi(q,\omega)]$ and then a normal momentum Fourier transform \cite{GenInt}. In the present case, where we are looking at only one momentum component, it is equally instructive to examine just the frequency transform, which yields the time dependence of a single spatial Fourier component of the density, $\chi(q,t)$. 

In Fig. 2 we plot the time evolution $\chi(q,t)$ determined from the curves shown in Fig. 1b,c.  The primary difference is a slowly varying background, which (sensibly) is a long-time consequence of the different low-frequency parts of the two spectra. However, the curves contain quantitative similarities. The detailed, early time dynamics--riding on top of the background--is identical between the two curves. Evidently, despite the anomalies at low energy, the fastest part of the dynamics is robust and independent of the way the anomaly is treated. We conclude that this aspect of the dynamics is reliable and characteristic of the true system.

So, is the resolution anomaly a problem? We conclude that the answer to this question is a matter of time scale. The fastest dynamics--i.e., that which occurs on time scales much less than the inverse of the energy width of the anomaly--are not affected and can be considered reliable. Intermediate times scales are somewhat reliable, apart from an uncontrolled, overall background. On time scales similar to the inverse width, the dynamics are totally aberrant.

\section{Summary}
In summary, synthesizing various elements from the literature of forty years ago, we have used contemporary notation to summarize a relationship between the cross section for HR-EELS, a correlation function, and the imaginary part of a response function. This relationship suggests an approach to performing real-time imaging of charge dynamics near the surface of a material. This approach would allow microscopic imaging of surface phonons and plasmons, and may find application in a wide variety of areas including strongly correlated electron systems, surface plasmonics, surface electronic structure determination, etc. We have found that the primary limitation on this technique is the division of the Bose factor, which is impeded in the low-frequency region by resolution effects. This creates anomalies in the time dynamics at long times. At time scales much shorter than the inverse width of the resolution function, however, the dynamics so achieved should be quite reliable.

\section{Acknowledgements}
This work was supported by the Office of Basic Energy Sciences, U.S. Department of Energy, Grant No. DE-FG02-06ER46285.

\bibliography{FDT-HREELS}{}
\bibliographystyle{elsarticle-num-names}
\end{document}